# A DEEP GENERATIVE ADVERSARIAL ARCHITECTURE FOR NETWORK-WIDE SPATIAL-TEMPORAL TRAFFIC STATE ESTIMATION

**Yunyi Liang, Ph.D. Candidate**
Key Laboratory of Road and Traffic Engineering of the Ministry of Education, Tongji University
4800 Cao' an Road, Shanghai, 201804 P. R. China
Tel: 86+15000190683; Email: liangyunyilyy@126.com

**Zhiyong Cui, Ph.D. Candidate**
Department of Civil and Environmental Engineering, University of Washington
Seattle, USA
Email: zhiyongc@uw.edu

**Yu Tian, Master student**
School of Economics and Management, Tongji University
Tongji Building A, 1 Zhangwu Road, Shanghai, 200092 P. R. China
Tel: 86+15000898586; Email: hellotianyu@126.com

**Huimiao Chen, Master student**
Department of Electrical Engineering, Tsinghua University
9-305, East-Main Building, Tsinghua University, Haidian, Beijing, P. R. China 100084
Tel: +86 18810307685; Email: chenhm15@mails.tsinghua.edu.cn

**Yinhai Wang, Professor, Ph.D. (Corresponding Author)**
Department of Civil and Environmental Engineering, University of Washington
Box 352700, Seattle, WA 98195-2700
Tel: (206) 616-269, Fax: (206) 543-1543, Email: yinhai@uw.edu

Word count: 5,730 words text + 7 tables/figures x 250 words (each) = 7,480 words

Submission Date: July 2017

## ABSTRACT

This study proposes a deep generative adversarial architecture (GAA) for network-wide spatial-temporal traffic state estimation. The GAA is able to combine traffic flow theory with neural networks and thus improve the accuracy of traffic state estimation. It consists of two Long Short-Term Memory Neural Networks (LSTM NNs) which capture correlation in time and space among traffic flow and traffic density. One of the LSTM NNs, called a discriminative network, aims to maximize the probability of assigning correct labels to both true traffic state matrices (i.e., traffic flow and traffic density within a given spatial-temporal area) and the traffic state matrices generated from the other neural network. The other LSTM NN, called a generative network, aims to generate traffic state matrices which maximize the probability that the discriminative network assigns true labels to them. The two LSTM NNs are trained simultaneously such that the trained generative network can generate traffic matrices similar to those in the training data set. Given a traffic state matrix with missing values, we use back-propagation on three defined loss functions to map the corrupted matrix to a latent space. The mapping vector is then passed through the pre-trained generative network to estimate the missing values of the corrupted matrix. The proposed GAA is compared with the existing Bayesian network approach on loop detector data collected from Seattle, Washington and that collected from San Diego, California. Experimental results indicate that the GAA can achieve higher accuracy in traffic state estimation than the Bayesian network approach.

## 1. INTRODUCTION

Traffic state usually refers to traffic flow, traffic density, traffic speed and travel time in the existing literature (*1*). Traffic state information plays an important role in relieving traffic congestion. With the aid of accurate traffic state information, administrators can effectively manage and control the transportation network, and travelers can better decide on their departure times and travel routes to make their trips more efficient.

With the rapid development of Intelligent Transportation Systems (ITS), enriched real-time traffic data can be collected by various types of sensors, such as loop detectors, Global Positioning Systems (GPS) and Remote Traffic Microwave Sensors (RTMS). Strategies for traffic management and control, developed based on the application of such data, are often not as effective as expected due to lag time (*2*). In addition, traffic data at some spatial and temporal positions may be missing due to sensor disruption. Therefore, traffic state estimation has become an important research topic in the transportation area and attracted much attention. Existing studies for traffic state estimation can be divided into parametric approaches, nonparametric approaches and hybrid integration methods (*3*). In parametric approaches, model structure is predetermined based on certain theoretical assumptions and model parameters are calibrated from empirical data. Typical methods of this type include cell-transmission-model-based methods (*4*) and dynamic-traffic-assignment-based methods (*5*). In nonparametric approaches, both model structure and parameters are not fixed. This type of approaches can be further divided into classical statistical models and Computational Intelligence (CI) models (*6*). Typical models of the former type include the Kalman Filter (*7, 8*) and support vector machine (*9, 10*). The artificial neural network (ANN) (*11-13*) is a typical model of the latter type. Deep learning technology refers to the neural networks with more than three layers. It is one of the most promising Artificial Intelligence (AI) technologies. Compared with traditional neural networks, they are able to effectively extract features from training data and play an important role in many areas, such as image recognition, speech recognition and image blending. In the field of transportation, deep learning models are mainly applied for prediction problems, such as traffic flow prediction (*14*) and speed prediction (*15*); to date, they have achieved promising results. Hybrid integration approaches combine the parametric approaches and nonparametric approaches. As pointed out by Vlahogianni et al. (*6*), CI models outperform classical statistical models in traffic state estimation because CI models have less a priori assumptions for input variables and are more capable of processing disrupted data.

Traffic flow theory, such as the cell transmission model (*4*), indicates that traffic flow parameters in a transportation network have correlation over both time and space. Incorporating traffic flow theory is beneficial for improving the accuracy of traffic state estimation. However, most of the CI models in the literature fail to do so due to their inflexible model structures.

Generative adversarial networks (GANs) are the frontier of deep learning technology. ANNs can be categorized into discriminative networks and generative networks. A trained discriminative network is able to predict output given input by learning the conditional probability distribution of output given input. While a trained generative network is able to generate samples having a similar distribution to the training samples by learning the joint distribution of the input and the output. Since Goodfellow et al. proposed GANs (*16*), GANs have become a very hot topic in the AI area. They have been applied in image inpainting, image blending, image translation and super-resolution. So far, the most striking successes in application of ANNs for traffic state estimation is based on discriminative networks. Having a flexible framework, GANs are able to incorporate the spatial-temporal relationship between traffic flow parameters in traffic state estimation and improve the accuracy of traffic state estimation.

The methodology presented this paper is inspired by GANs used for image inpainting (*17*). The principle of image inpainting using GANs mainly depends on two types of information: 1) the prior information in the corrupted image, and 2) the experience from the training data set. The former makes the recovered image logical while the latter makes the recovered image realistic. The methodology proposed by this study is analogous to the principle of image inpainting. We regard traffic state within the time and space plane as an image. Then we estimate traffic state using three types of information. The first type of information is initial spatial-temporal traffic state. The second type of information is traffic state correlation in time and space learned from training data. The last type of information is the conservation law of traffic flow which in theory describes the spatial-temporal relationship between traffic flow and traffic density. By combining these pieces of information, we are able to estimate spatial-temporal traffic state with a high level of accuracy.

The remainder of this article is organized as follows. Section 2 briefly introduces GANs and LSTM NNs. The generative adversarial architecture (GAA) for spatial-temporal traffic state estimation is proposed in Section 3. In Section 4, the proposed GAA is compared with the existing Bayesian network approach on two data sets: loop detector data collected from Seattle, Washington and that collected from San Diego, California. Section 5 concludes this study and discusses future work.

## 2 GANS AND LSTM NNS
### 2.1 Generative adversarial networks (GANs)
GANs are frameworks for training generative models. GANs consist of two neural networks, a generative neural network $G$, and a discriminative neural network $D$. $G$ maps a random vector $\mathbf{z}$, sampled from a prior distribution $p_{\mathbf{z}}$, to the sample space. $D$ maps an input sample to a likelihood. $G$ aims to generate realistic samples to get true labels from $D$. While $D$ plays an adversarial role to discriminate between the sample generated from $G$ and the sample from the training data. The training procedure for GANs corresponds to a minimax two-player game. It converges when Nash equilibrium is achieved. Thus, the two neural networks are simultaneously trained by optimizing the following loss function:

$$\min_{G} \max_{D} V(G, D) = E_{\mathbf{h} \sim p_{data}(\mathbf{h})}[\log(D(\mathbf{h}))] + E_{\mathbf{z} \sim p_{\mathbf{z}}(\mathbf{z})}[\log(1 - D(G(\mathbf{z})))] \tag{1}$$

where $\mathbf{h}$ is sampled from the training data distribution $p_{data}$, $\mathbf{z}$ is a random vector sampled from a prior distribution $p_{\mathbf{z}}$.

The convergence to the global optimality of the training procedure has been proved in (*16*). After convergence, $G$ can produce a high-quality sample similar to the training data distribution from the latent space and $D$ is not able to tell generated samples from training data. That said, $D$ is able to reject samples that are too fake or dissimilar. The principle of GANs is illustrated in Figure 1(a). Note that the two neural networks can be of any type. In this study, we use two Long Short Term Memory Neural Networks (LSTM NNs) to capture the spatial-temporal correlation of traffic state in a transportation network.

## 2.2 Long Short Term Memory Neural Networks (LSTM NNs)
Recurrent neural networks (RNNs) are a class of artificial neural networks. They have the capability of modeling nonlinear time series problems by feeding back the current state of each hidden layer to itself along with inputs at the next time step.

The principle of a basic RNN is illustrated in Figure 1(b). Where $\mathbf{x}_t$, $\mathbf{o}_t$ and $\mathbf{h}_t$ are the input to the RNN at time step $t$, the output of the RNN at time step $t$ and the hidden layer state of the RNN at time step $t$, respectively. For further details about RNNs, readers are referred to (*15*). Despite the superior capability of RNNs in modeling nonlinear time series problems, when modeling time series problem with long time lags, the traditional RNNs encounter the vanishing gradient problem. The vanishing gradient problem in the context of training RNNs means that the update of weight matrix $W$ is nearly unaffected by the hidden layer state $h$ in time steps long before the current one.

To overcome the vanishing gradient problem of traditional RNNs, LSTM NNs were initially introduced by (*18*). The primary objectives of LSTM NNs are to model long-term dependencies and determine the optimal time lag for time series problems. As pointed out in (*15*), these two features are desirable for traffic state estimation. A basic LSTM NN is composed of one input layer, one hidden layer and one output layer. The hidden layer is the core of an LSTM NN. Compared with that of a traditional RNN, the hidden layer of an LSTM NN has one more state, the cell state $\mathbf{c}$, for long term memory storage. In an LSTM NN, the cell state $\mathbf{c}$ is controlled by two gates, a forget gate and an input gate. The former determines how much of the cell state $c_{t-1}$ at the last time step is kept at the current cell state $c_t$. The latter determines how much of the current input $x_t$ of the LSTM NN is kept at the current cell state $c_t$. The output of the forget gate $\mathbf{f_t}$ and the input gate $\mathbf{i_t}$ can be calculated by Eq. (2) and Eq. (3), respectively

$$\mathbf{f_t} = \sigma(\mathbf{W_f} \cdot [\mathbf{h_{t-1}}, \ \mathbf{x_t}] + \mathbf{b_f}) \tag{2}$$

$$\mathbf{i_t} = \sigma(\mathbf{W_i} \cdot [\mathbf{h_{t-1}}, \ \mathbf{x_t}] + \mathbf{b_i}) \tag{3}$$

where $\mathbf{W_f}$ and $\mathbf{W_i}$ are the weight matrices of forget gate and input gate, respectively; $[\mathbf{h_{t-1}}, \ \mathbf{x_t}]$ is the new vector formed by connecting $\mathbf{h_{t-1}}$ and $\mathbf{x_t}$; $\mathbf{b_f}$ and $\mathbf{b_i}$ are the bias vectors of forget gate and input gate, respectively. $\sigma(\cdot)$ denotes the standard logistic sigmoid function defined in Eq. (4)

$$\sigma(x) = \frac{1}{1 + e^{-x}} \tag{4}$$

The current memory $\mathbf{Q_t}$ can be expressed by

$$\mathbf{Q_t} = \tanh(\mathbf{W}_c \cdot [\mathbf{h_{t-1}}, \ \mathbf{x_t}] + \mathbf{b_c}) \tag{5}$$

where $\mathbf{W}_c$ and $\mathbf{b_c}$ are the corresponding weight matrix and bias vector, respectively. Function $\tanh(\cdot)$ is defined as

$$\tanh(x) = \frac{e^x - e^{-x}}{e^x + e^{-x}} \tag{6}$$

Then the current cell state $\mathbf{c_t}$ can be calculated by

$$\mathbf{c_t} = \mathbf{f_t} \odot \mathbf{c_{t-1}} + \mathbf{i_t} \odot \mathbf{Q_t} \tag{7}$$

where $\odot$ represents the scalar product of two vectors. Using Eq. (7), we combine long term memory $\mathbf{c_{t-1}}$ and current memory $\mathbf{Q_t}$ to determine the current cell state $\mathbf{c_t}$ in an LSTM NN. Therefore, the LSTM NN is able to store the information long before it is due for processing at the forget gate, while it is also able to incorporate the current information into memory.

The output gate calculated by Eq. (8) describes the effect of long term memory on the current output:

$$\mathbf{o_t} = \sigma(\mathbf{W_o} \cdot [\mathbf{h_t}, \ \mathbf{x_t}] + \mathbf{b_o}) \tag{8}$$

where $\mathbf{W}_c$ and $\mathbf{b}_c$ are the weight matrix and bias vector of the output gate, respectively.

The final output of an LSTM cell can be determined by the output gate $\mathbf{o_t}$ and the current cell state $\mathbf{c_t}$

$$\mathbf{h_t} = \mathbf{o_t} \odot \tanh(\mathbf{c_t}) \tag{9}$$

The principle of a basic LSTM NN is illustrated in Figure 1(c).

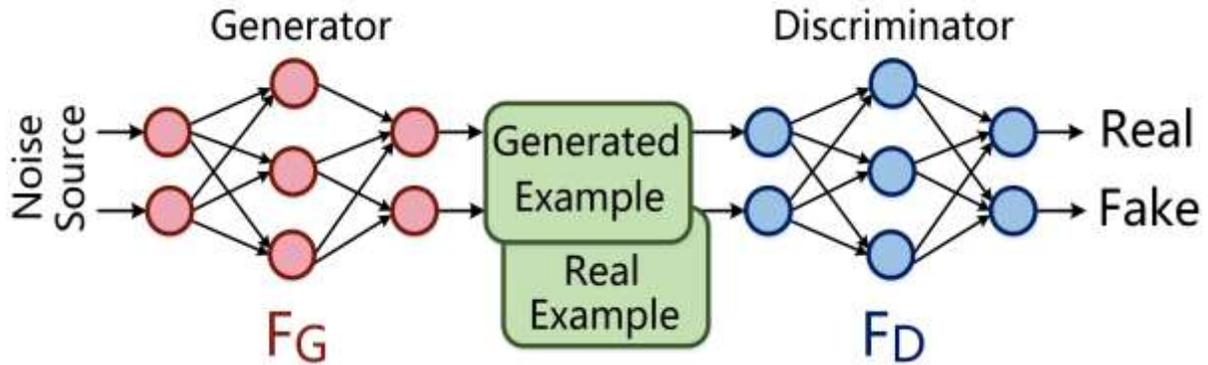

FIGURE 1(a) The principle of generative adversarial networks (GANs).

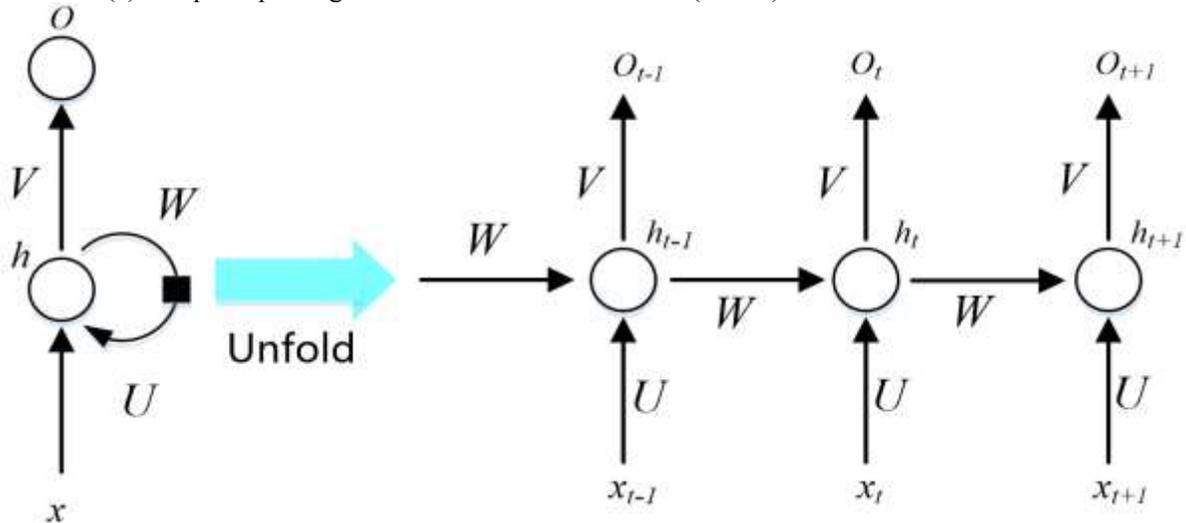

FIGURE 1(b) The principle of a basic recurrent neural network (RNN).

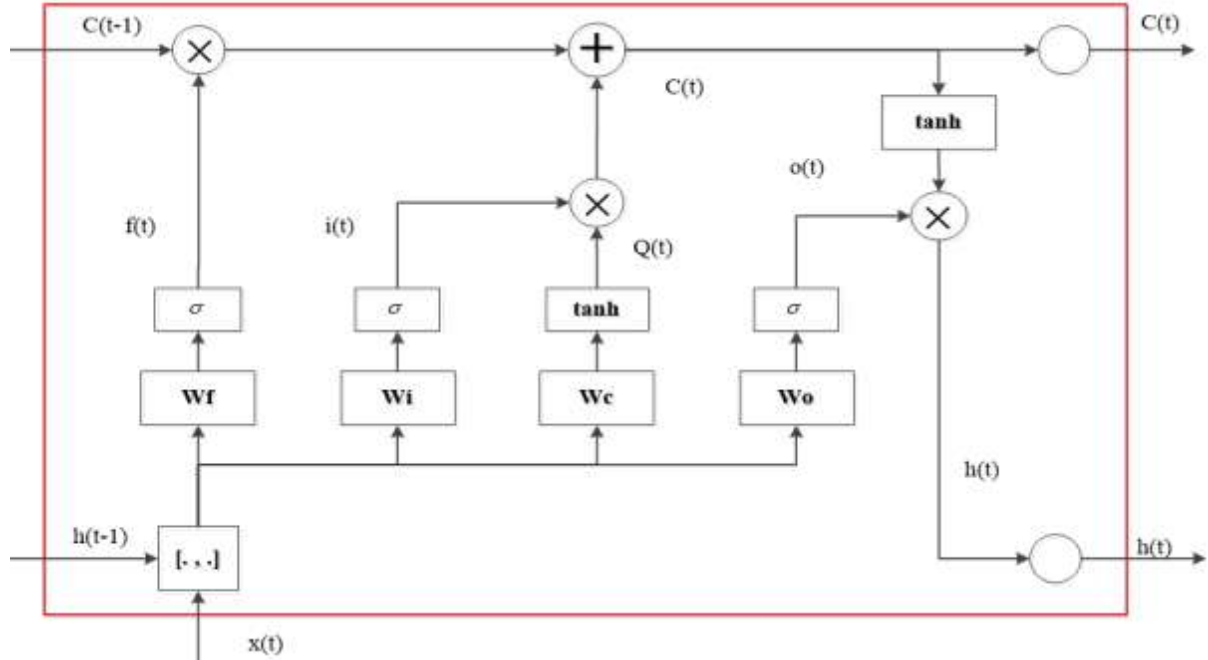

FIGURE 1(c) The principle of a basic long short term memory neural network (LSTM NN).
**FIGURE 1 The principle of GANs, a basic RNN and a basic LSTM NN.**

## 3   GENERATIVE ADVERSARIAL ARCHITECTURE FOR SPATIAL-TEMPORAL TRAFFIC STATE ESTIMATION

### 3.1 Motivation and problem description

Traffic flow theory, such as the cell transmission model (*4*), indicates that traffic flow, traffic density and traffic velocity in a transportation network are correlated. Futher, these correlations exist in both time and space. Incorporating these spatial-temporal correlations can improve the accuracy of traffic state estimation. For example, Ma et al. (*15*) chose traffic flow and traffic velocity as inputs for LSTM NNs to improve traffic velocity prediction. Spatial-temporal correlations among traffic flow parameters are incorporated in traffic state estimation in (*2*) (*9*) (*20*). However, in those studies, spatial-temporal correlations among traffic flow parameters are only mined from data and no specific theoretical correlations are defined due to their relatively fixed model structures. As shown in Section 4 later, GANs are able to incorporate correlations among traffic flow parameters from both the data and theory due to their flexible framework.

The traffic state estimation problem addressed in this study can be described as follows. We choose traffic flow and traffic density as inputs to our proposed model. As traffic flow and traffic density evolve with time and space, both traffic flow and traffic density within a given spatial-temporal area can be represented as a two-dimensional matrix.  We assume the number of time intervals for the study period is $n$, and that the road segment can be divided into $m$ discrete cells. The $n*(m+1)$ traffic flow matrix is denoted as $F$ with each element $q(t,l)$ representing the traffic flow at location $l$ at time $t$. The $n*m$ traffic density matrix is denoted as $K$ with each element $k(t,s)$ representing the traffic density in cell $s$ at time $t$. The problem addressed in this study is how to estimate or predict the missing values in $F$ and $K$ given known elements in each matrix. Note that the proposed model can be viewed as a deep-learning-based and data-driven cell transmission model if the known elements are boundary conditions. The problem description can be explained by Figure 2. Note that we do not add traffic velocity as an input in this study. This is mainly because traffic velocity $v$ can be determined by traffic flow and traffic density using $v = q / k$ and thus it makes little contribution to traffic state estimation.

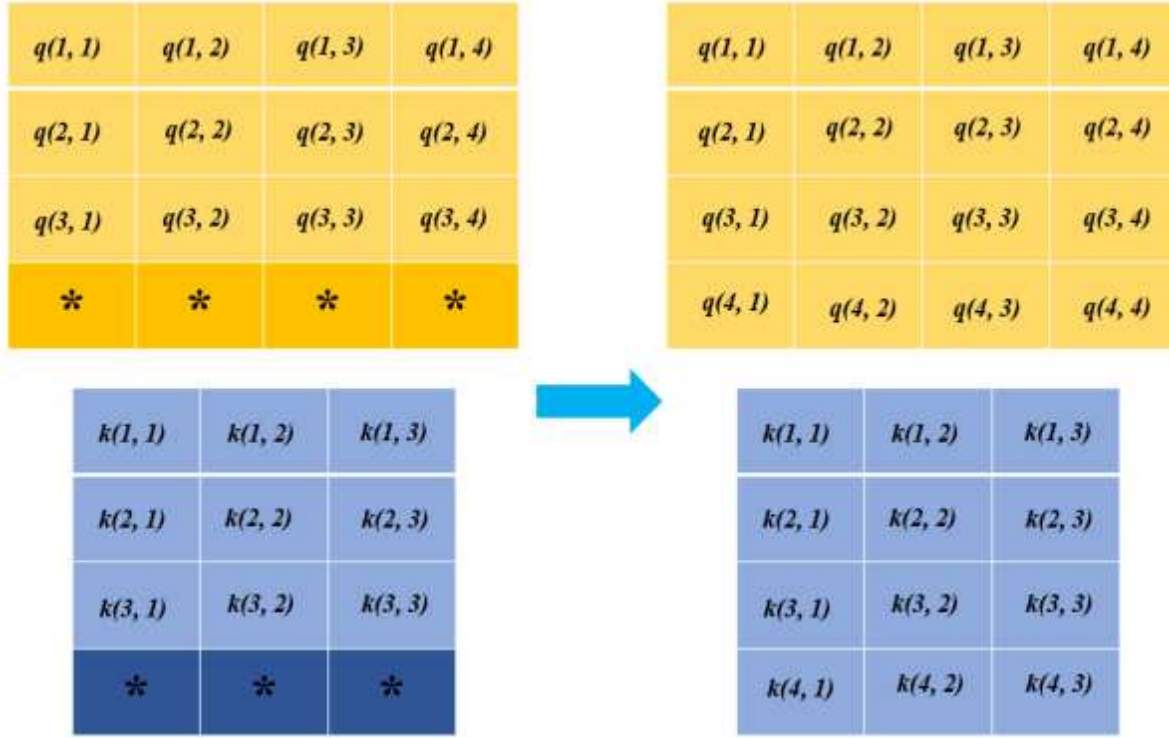

**Figure 2 Description of the traffic state estimation problem addressed in this study**

### 3.2 Network architecture

The generative adversarial architecture proposed in this study consists of two LSTM NNs. In our generative adversarial architecture, the generative LSTM NN randomly generates traffic state matrices. Another LSTM NN, the discriminative network, estimates the probability that a traffic state matrix comes from training data rather than the generative LSTM NN. We train the discriminative LSTM NN to maximize the probability of assigning correct labels to both traffic state matrices from training data and traffic state matrices from the generative LSTM NN. The generative LSTM NN is trained to generate realistic traffic state matrices which maximize the probability that the discriminative LSTM NN assigns true labels to them. The discriminative LSTM NN has four layers: one input layer, two hidden layers and one output layer (see Figure 3(a)). The input of the discriminative LSTM NN is a sequence with time steps. At each time step, the inputs are the traffic flow at each location and the traffic density at each cell. The first hidden layer contains a LSTM cell as described in Section 2.2. The output layer maps the output sequence of the second hidden layer into a value between 0 and 1 such that it is able to calculate the probability that the input traffic state matrix is from training data. The sigmoid function defined by Eq. (4) is used in the output layer.

The input of the generative LSTM NN is a sequence randomly sampled from a uniform distribution following (*16*). Note that other kinds of distribution can be used to replace the uniform distribution. Its output is a traffic state matrix having the same dimension as the output of the second hidden layer of the discriminative LSTM NN. The model structure of the generative LSTM NN is shown in Figure 3(b).

The two LSTM NNs are trained simulatenously using the GANs framework described in Section 2.1. The training procedure using minibatch stochastic gradient descent is illustrated in Figure 3(c).

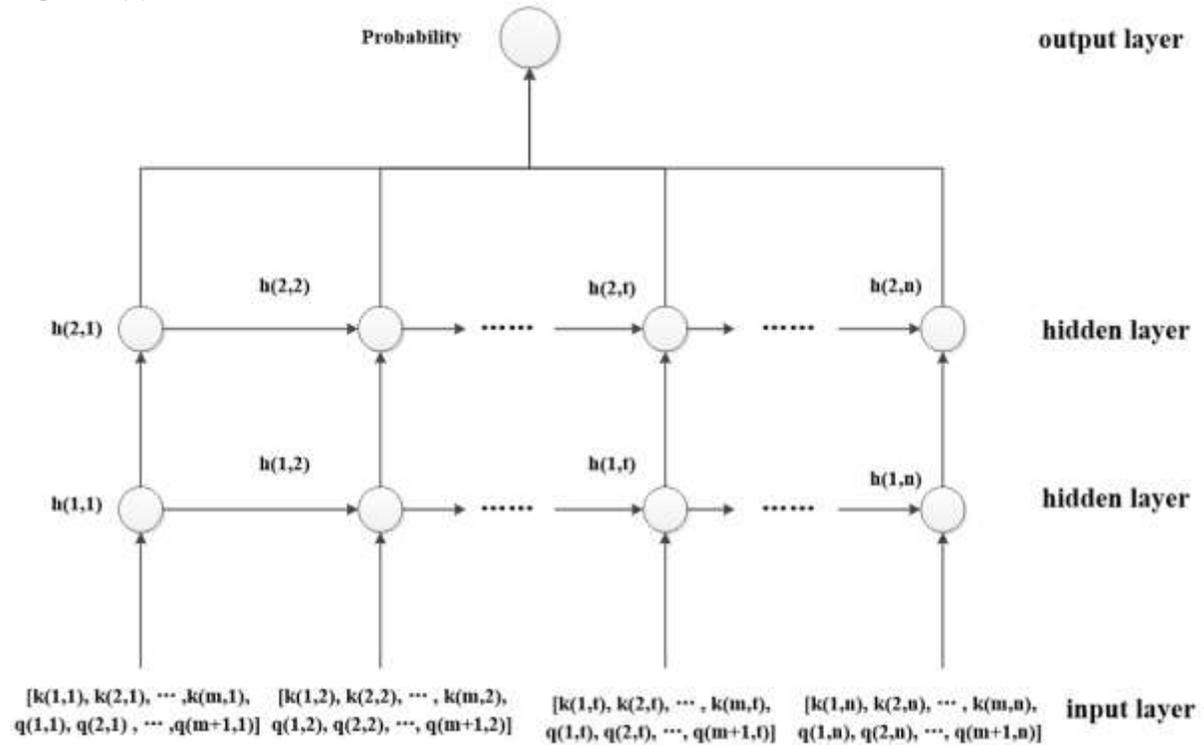

FIGURE 3(a) Model structure of the discriminative Long Short Term Memory Neural Network.

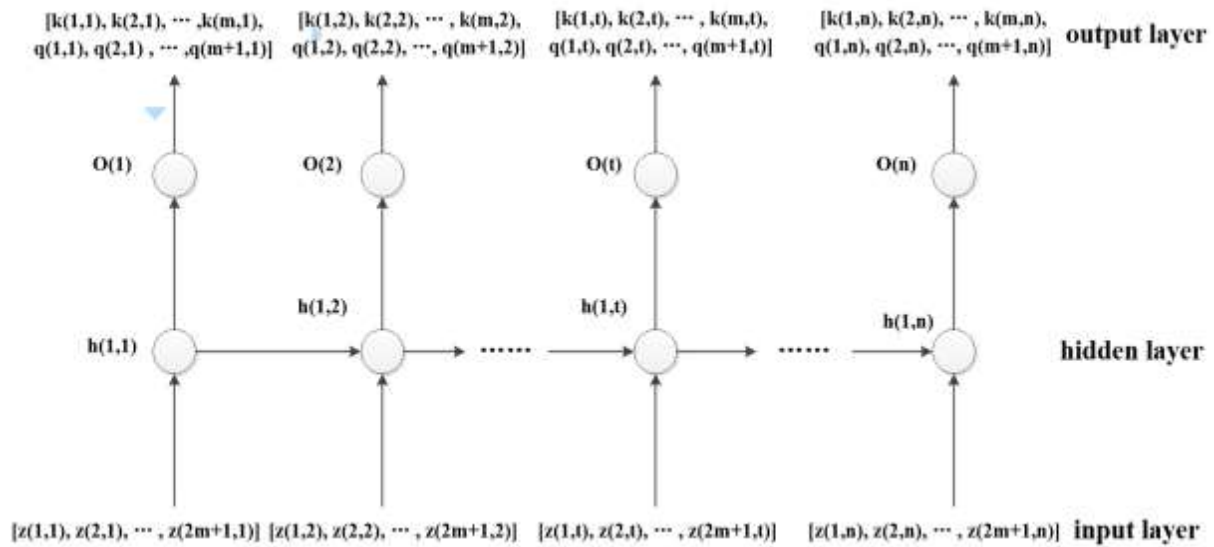

FIGURE 3(b) Model structure of the generative Long Short Term Memory Neural Network.

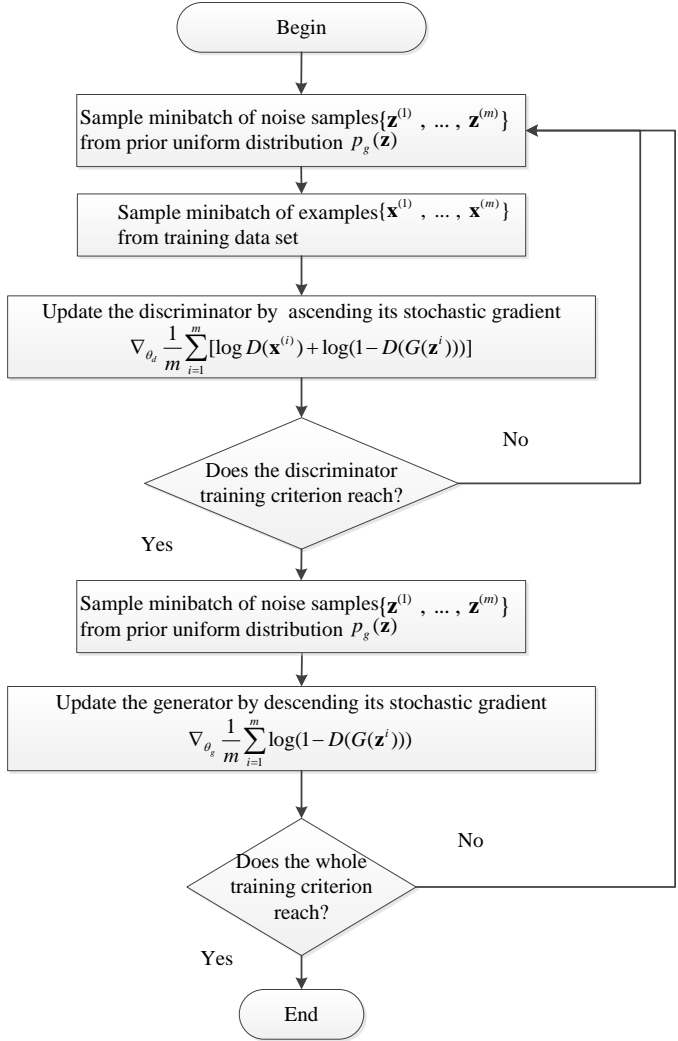

FIGURE 3(c) Minibatch stochastic gradient descent training of generative adversarial networks.

**FIGURE 3 Model structure of the proposed generative adversarial architecture (GAA).**

Within the GANs framework, each of the two LSTM NNs are trained by the algorithm proposed in *(19)*.The steps of the training algorithm for each LSTM NN are described as follows.

Step 1: forward calculate the values of $\mathbf{f}_t$, $\mathbf{i}_t$, $\mathbf{c}_t$, $\mathbf{o}_t$ and $\mathbf{h}_t$ using Eq. (2) to Eq. (9).

Step 2: Backward calculate the error in LSTM NN training. The backward calculation of the error includes the error that propagates backward over time (i.e., calculate the error at each previous time step from the current time step) and the error that propagates backward to previous layers from the current layer. The error at time step $p$ backward propagating from time step $t$ can be calculated by

$$\delta_p^T = \prod_{j=p}^{t-1} \delta_{o,t}^T \mathbf{W}_{oh} + \delta_{f,t}^T \mathbf{W}_{fh} + \delta_{i,t}^T \mathbf{W}_{ih} + \delta_{Q,t}^T \mathbf{W}_{ch} \tag{10}$$

where

$$\delta_{o,t}^T = \delta_t^T \circ \tanh(\mathbf{c}_t) \circ \mathbf{o}_t \circ (1 - \mathbf{o}_t) \tag{11}$$

$$\delta_{f,t}^T = \delta_t^T \circ \mathbf{o}_t \circ (1 - \tanh(\mathbf{c}_t)^2) \circ \mathbf{c}_{t-1} \circ \mathbf{f}_t \circ (1 - \mathbf{f}_t) \tag{12}$$

$$\delta_{i,t}^T = \delta_t^T \circ \mathbf{o}_t \circ (1 - \tanh(\mathbf{c}_t)^2) \circ \mathbf{Q}_{t-1} \circ \mathbf{i}_t \circ (1 - \mathbf{i}_t) \tag{13}$$

$$\delta_{i,t}^T = \delta_t^T \circ \mathbf{o}_t \circ (1 - \tanh(\mathbf{c}_t)^2) \circ \mathbf{i}_t \circ (1 - \mathbf{Q}_t^2) \tag{14}$$

The error at layer $(l-1)$ backward propagating from layer $l$ can be calculated by

$$\frac{\partial E}{\partial \mathbf{net}_t^{l-1}} = (\delta_{o,t}^T \mathbf{W}_{oh} + \delta_{f,t}^T \mathbf{W}_{fh} + \delta_{i,t}^T \mathbf{W}_{ih} + \delta_{Q,t}^T \mathbf{W}_{ch}) \circ z^{'}(\mathbf{net}_t^{l-1}) \tag{15}$$

where $\mathbf{net}_t^{l-1}$ is the input at layer $(l-1)$ at time step $t$, $z$ is the activation function at layer $(l-1)$.

Step 3: calculate the gradient of weight matrices and the bias vectors according to their corresponding errors using Eq. (16) to Eq. (27).

$$\frac{\partial E}{\partial \mathbf{W}_{oh}} = \sum_{j=1}^{t} \delta_{o,j}^T \mathbf{h}_{j-1}^T \tag{16}$$

$$\frac{\partial E}{\partial \mathbf{W}_{fh}} = \sum_{j=1}^{t} \delta_{f,j}^T \mathbf{h}_{j-1}^T \tag{17}$$

$$\frac{\partial E}{\partial \mathbf{W}_{ih}} = \sum_{j=1}^{t} \delta_{i,j}^T \mathbf{h}_{j-1}^T \tag{18}$$

$$\frac{\partial E}{\partial \mathbf{W}_{ch}} = \sum_{j=1}^{t} \delta_{Q,j}^T \mathbf{h}_{j-1}^T \tag{19}$$

$$\frac{\partial E}{\partial \mathbf{b}_o} = \sum_{j=1}^{t} \delta_{o,j} \tag{20}$$

$$\frac{\partial E}{\partial \mathbf{b}_i} = \sum_{j=1}^{t} \delta_{i,j} \tag{21}$$

$$\frac{\partial E}{\partial \mathbf{b}_f} = \sum_{j=1}^{t} \delta_{f,j} \tag{22}$$

$$\frac{\partial E}{\partial \mathbf{b}_c} = \sum_{j=1}^{t} \delta_{Q,j} \tag{23}$$

$$\frac{\partial E}{\partial \mathbf{W}_{ox}} = \delta_{o,t} \mathbf{x}_t^T \tag{24}$$

$$\frac{\partial E}{\partial \mathbf{W}_{fx}} = \delta_{f,t} \mathbf{x}_t^T \tag{25}$$

$$\frac{\partial E}{\partial \mathbf{W}_{ix}} = \delta_{i,t} \mathbf{x}_t^T \tag{26}$$

$$\frac{\partial E}{\partial \mathbf{W}_{cx}} = \delta_{Q,t} \mathbf{x}_t^T \tag{27}$$

where $\mathbf{W}_{oh}$, $\mathbf{W}_{fh}$, $\mathbf{W}_{ih}$, $\mathbf{W}_{ch}$, $\mathbf{W}_{ox}$, $\mathbf{W}_{fx}$, $\mathbf{W}_{ix}$ and $\mathbf{W}_{cx}$ satisfy

$$[\mathbf{W}_o]\begin{bmatrix}\mathbf{h}_{t-1}\\\mathbf{x}_t\end{bmatrix} = [\mathbf{W}_{oh}\ \mathbf{W}_{ox}]\begin{bmatrix}\mathbf{h}_{t-1}\\\mathbf{x}_t\end{bmatrix} = \mathbf{W}_{oh}\mathbf{h}_{t-1} + \mathbf{W}_{ox}\mathbf{x}_t \tag{28}$$

$$[\mathbf{W}_f]\begin{bmatrix}\mathbf{h}_{t-1}\\ \mathbf{x}_t\end{bmatrix} = [\mathbf{W}_{fh}\ \mathbf{W}_{fx}]\begin{bmatrix}\mathbf{h}_{t-1}\\ \mathbf{x}_t\end{bmatrix} = \mathbf{W}_{fh}\mathbf{h}_{t-1} + \mathbf{W}_{fx}\mathbf{x}_t \tag{29}$$

$$[\mathbf{W}_i]\begin{bmatrix}\mathbf{h}_{t-1}\\ \mathbf{x}_t\end{bmatrix} = [\mathbf{W}_{ih}\ \mathbf{W}_{ix}]\begin{bmatrix}\mathbf{h}_{t-1}\\ \mathbf{x}_t\end{bmatrix} = \mathbf{W}_{ih}\mathbf{h}_{t-1} + \mathbf{W}_{ix}\mathbf{x}_t \tag{30}$$

$$[\mathbf{W}_c]\begin{bmatrix}\mathbf{h}_{t-1}\\ \mathbf{x}_t\end{bmatrix} = [\mathbf{W}_{ch}\ \mathbf{W}_{cx}]\begin{bmatrix}\mathbf{h}_{t-1}\\ \mathbf{x}_t\end{bmatrix} = \mathbf{W}_{ch}\mathbf{h}_{t-1} + \mathbf{W}_{cx}\mathbf{x}_t \tag{31}$$

### 3.3 Traffic state estimation with pre-trained GANs

Our methodology for traffic state estimation utilizes the $G$ and $D$ networks, pre-trained with uncorrupted data, to reconstruct traffic state. Denote the corrupted traffic state matrix as $\mathbf{y}$. We do not use $D$ to update $\mathbf{y}$ by maximizing $D(\mathbf{y})$. Similar to the image inpainting in (*17*), maximizing $D(\mathbf{y})$ does not lead to the desired reconstruction. This is mainly because $\mathbf{y}$ is neither on the $p_{data}$ manifold, nor on the $G$ manifold and the corrupted data is not drawn from those distributions. Therefore, we consider using both $G$ and $D$ for reconstruction. To quantify the 'closest' mapping from $\mathbf{y}$ to the reconstruction, we define three loss functions: a contextual loss, a perceptual loss and a conservative loss.

*Contextual loss*

We need to incorporate the information from the known information of the given corrupted traffic state matrix into the traffic state estimation. The contextual loss is used to measure the context similarity between the reconstructed traffic state matrix and the uncorrupted portion. The contextual loss is defined as

$$L_{contextual}(\mathbf{z}) = \left\| M \odot G(\mathbf{z}) - M \odot \mathbf{y} \right\|_1 \tag{32}$$

where $M$ is a binary matrix having the same dimension as the output traffic matrices of $G$, denoting the mask of the corrupted traffic matrix. Elements equal to one in $M$ indicate the corresponding element in the corrupted traffic matrix is uncorrupted. Elements equal to zero in $M$ indicate the corresponding element in the corrupted traffic matrix is missing. $\odot$ denotes the element-wise product operation. The corrupted portion, i.e., $(1-M) \odot \mathbf{y}$ is not used in the loss. The choice of the $\ell_1$-norm is empirical. As shown in (*17*), in image inpainting, images recovered with the $\ell_1$-norm tend to be sharper and of higher quality compared to the ones reconstructed with the $\ell_2$-norm.

*Perceptual loss*

The perceptual loss encourages the reconstructed traffic state matrix to be similar to the samples drawn from the training set. This is achieved by updating $\mathbf{z}$ to fool $D$. As a result, $D$ will predict $G(\mathbf{z})$ to be from the training data with a high probability. We use the same loss for fooling $D$ as in GANs

$$L_{perceptual}(\mathbf{z}) = \log(1 - D(G(\mathbf{z}))) \tag{33}$$

Without $L_{perceptual}$, some reconstructed traffic state matrices tend to be unrealistic. We illustrate this by showing the examples where we optimized with and without $L_{perceptual}$ in Section 4.

*Conservative loss*

According to the conservation law of traffic flow (*4*), the spatial-temporal relationship between traffic volume and traffic density can be described as

$$k(t+1,s) = k(t,s) + \frac{\Delta t}{\Delta x_s}(q_{in}(t,s) - q_{out}(t,s)) \tag{34}$$

where $k(t,s)$ is the vehicle density of cell $s$ at time index $t$, $q_{in}(t,s)$ and $q_{out}(t,s)$ are the total flows (in vehicles per unit time) entering and leaving cell $s$ during the time interval $[t \cdot \Delta t,\ (t+1) \cdot \Delta t)$ respectively. $\Delta t$ is the sampling duration, and $\Delta x_s$ is the length of cell $s$. In the context of traffic state estimation using GAA, the conservative loss is defined as

$$K_{G(\mathbf{z})}(t+1,s) = K_{G(\mathbf{z})}(t,s) + \frac{\Delta t}{\Delta x_s}(F_{G(\mathbf{z})}(t,s) - F_{G(\mathbf{z})}(t,s+1)) \tag{35}$$

where $K_{G(\mathbf{z})}$ and $F_{G(\mathbf{z})}$ are the traffic density matrix and the traffic flow matrix generated by the pretrained $G$, respectively. Eq. (35) describes the theoretical spatial-temporal correlation among traffic flow and traffic density. Thus, it helps to improve the accuracy of traffic state estimation.

### 3.4 Traffic state estimation
With the defined loss functions, the traffic state matrix with missing values can be mapped to the closest $\mathbf{z}$ in the latent representation space. $\mathbf{z}$ is updated using back-propagation with the total loss

$$\hat{\mathbf{z}} = \arg\min_{\mathbf{z}}(L_{contextual}(\mathbf{z}) + \lambda_p L_{perceptual}(\mathbf{z}) + \lambda_c L_{conservative}(\mathbf{z})) \tag{36}$$

where $\lambda_p$ and $\lambda_c$ are weighting parameters. In practice, the weighting parameters have to be relatively small to ensure the traffic state estimation accuracy. After finding $\hat{\mathbf{z}}$, the estimated traffic state can be obtained by:

$$\mathbf{x}_{reconstructed} = \mathbf{M} \odot \mathbf{y} + (1-\mathbf{M}) \odot G(\hat{\mathbf{z}}) \tag{37}$$

### 4 EXPERIMENTS
The proposed generative adversarial architecture for spatial-temporal traffic state estimation is evaluated on two data sets: loop data collected from a segment of the I-5 freeway in Seattle, Washington (see Figure 4(a)) and loop data from a segment of the CA-52 state highway in San Diego, California (see Figure 4(b)). The segment of interest of the I-5 freeway has six loop detectors installed along it. The data is collected every 20 seconds from January 1st, 2015 to December 31st, 2015 and is stored in the Digital Roadway Interactive Visualization and Evaulation Network (DRIVE Net) system operated by the Smart Transportation Application and Research Laboratory (STAR Lab) at the University of Washington. The segement of interest along the CA-52 state highway has four loop detectors installed along it. Here, the data is collected every 30 seconds from January 1st, 2016 to December 31st, 2016 and is stored in the Performance Measurement System operated by the California Department of Transportation.

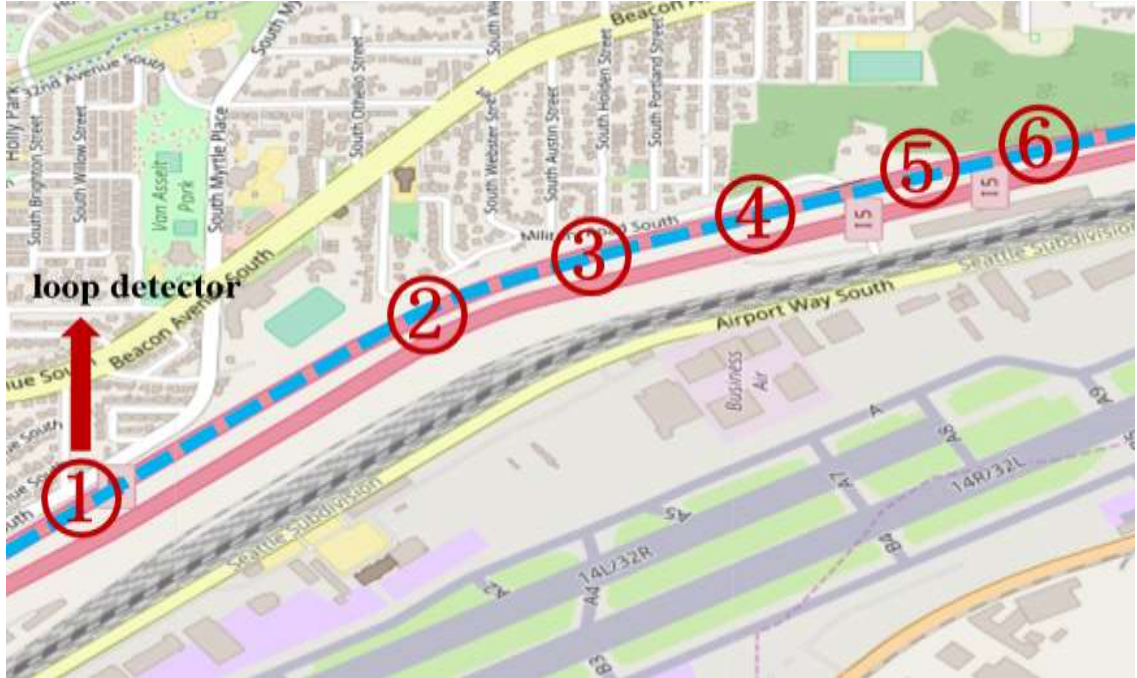

FIGURE 4(a) The I-5 freeway segment in Seattle, Washington.

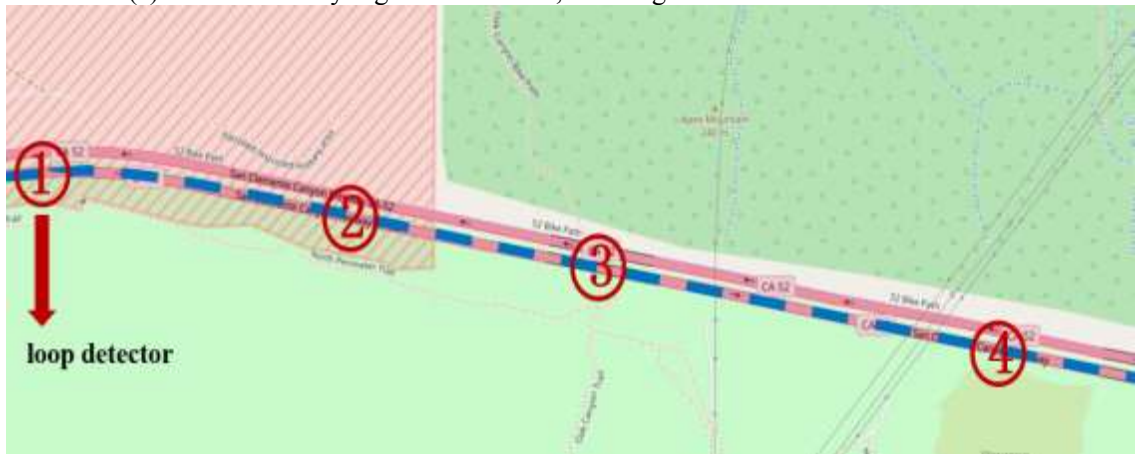

FIGURE 4(b) The CA-52 state highway segment in San Diego, California.

**FIGURE 4 The sites for data collection.**

Each training data record in the two experiments consists of data across one hour from the six loop detectors and the four loop detectors, respectively. This indicates that the dimension of each training traffic state matrix is 12*11 and 12*7 in the two experiments, respectively. In the two evaluations, we randomly select two-third of the data for training and the rest is used for validation.

Four types of GAAs are evaluated in the two experiments. These types include a GAA without perceptual loss and conservative loss (GAA model 1), a GAA without perceptual loss (GAA model 2), a GAA without conservative loss (GAA model 3) and a GAA with perceptual loss and conservative loss (GAA model 4). The four types of GAAs are compared with the existing Bayesian network approach proposed by (*20*) in each experiment. Similar to our proposed GAA models, the Bayesian network approach considers spatial-temporal correlation among traffic flow parameters in traffic state estimation. Therefore, the Bayesian network approach is chosen for comparison in this study.

Discriminative LSTM NNs and generative LSTM NNs in the four GAAs respectively have the same topology as shown in Figure 3(a) and Figure 3(b) and they have 16 hidden neurons in the hidden layer.

Each record in the two data sets is a pair of average traffic flow and traffic density measurements computed over one hour. We used the data in the previous half hour (i.e., 6*5 minutes) to predict traffic flow and traffic density in the next half hour (i.e., the next 6*5minutes). Note that this time lag is not limited to one hour as the LSTM NN is capable of capturing long time dependency among traffic flow parameters. Figure 5 and Figure 6 compare the traffic state estimated by the five models on the two data sets, respectively. Based on the data collected for the I-5 freeway segment, subfigures 5(a)-5(f) compare the traffic density estimated by the five models from 7:00AM to 7:00PM on October 10[th] between each pair of loop detectors, respectively. Subfigures 5(g)-5(k) compare the traffic flow estimated by the five models in the same way. As opposed to the traffic density defined between each pair of loop detectors, traffic flow is defined on each loop detector. Figure 6 presents results in the analogous and aforementioned manner, but based on the data collected along the CA-52 state highway.

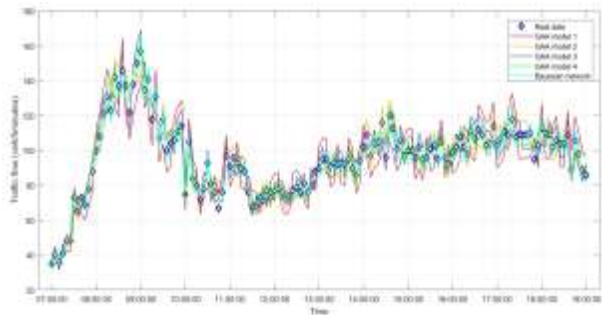

FIGURE 5(a) Predicted traffic flow for loop detector 1.

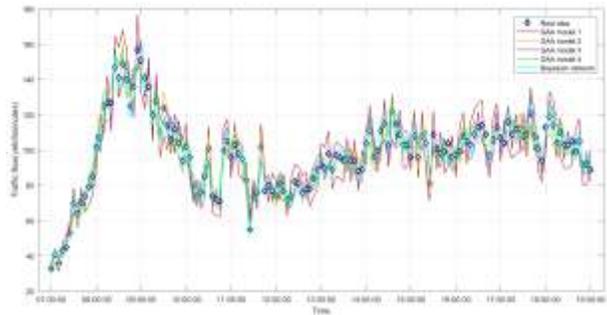

FIGURE 5(b) Predicted traffic flow for loop detector 2.

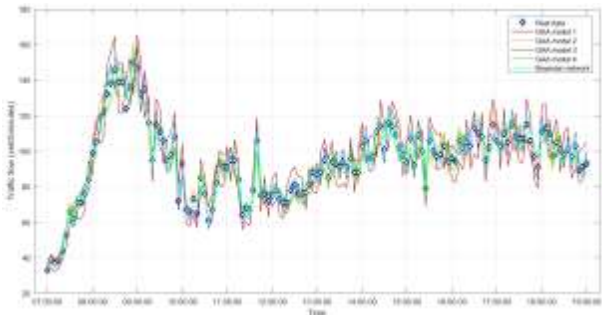

FIGURE 5(c) Predicted traffic flow for loop detector 3.

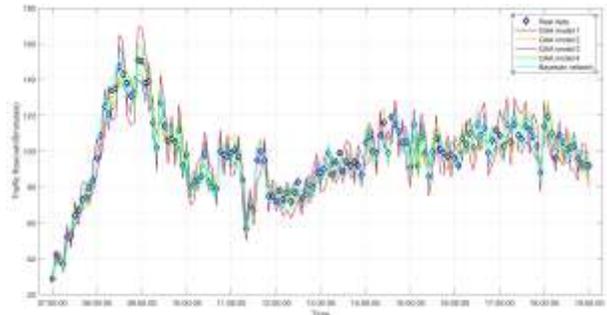

FIGURE 5(d) Predicted traffic flow for loop detector 4.

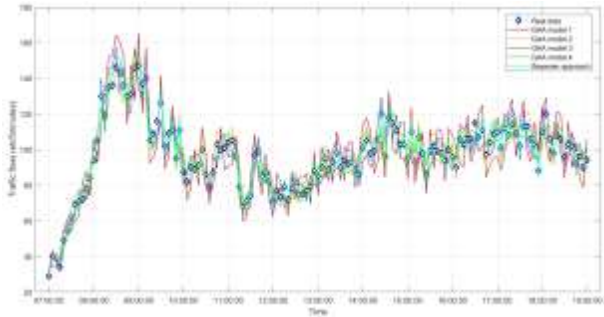 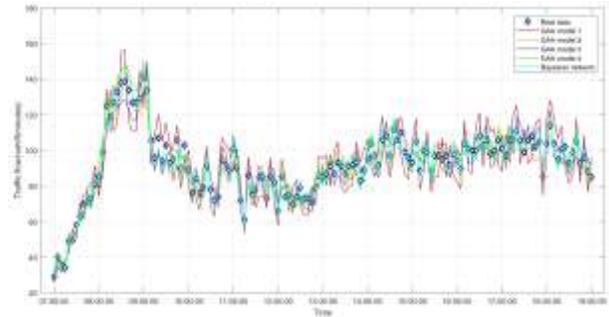

FIGURE 5(e) Predicted traffic flow for loop detector 5.

FIGURE 5(f) Predicted traffic flow for loop detector 6.

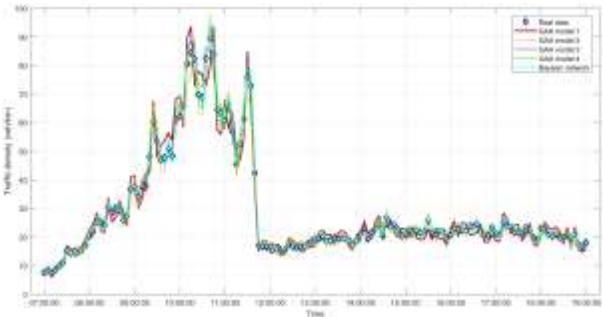 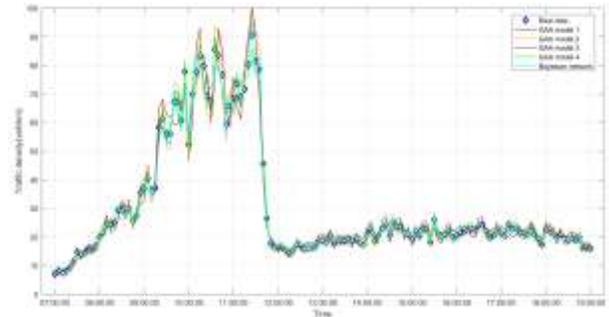

FIGURE 5(g) Predicted traffic density between loop detector 1 and 2.

FIGURE 5(h) Predicted traffic density between loop detector 2 and 3.

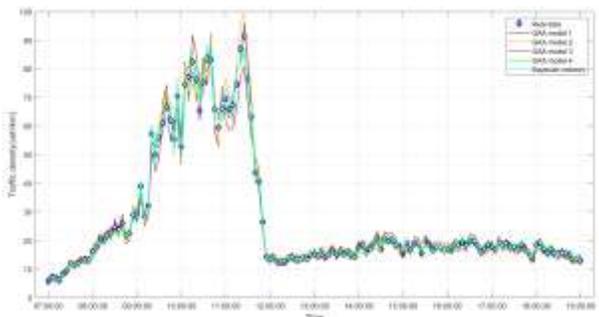 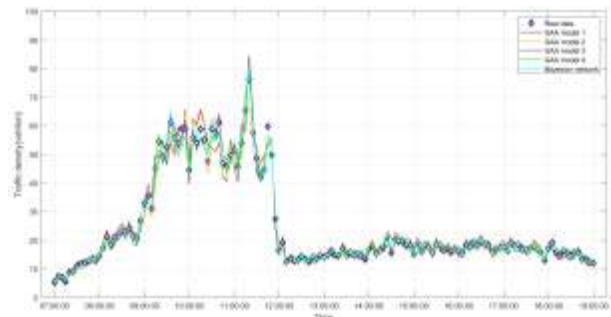

FIGURE 5(i) Predicted traffic density between loop detector 3 and 4.

FIGURE 5(j) Predicted traffic density between loop detector 4 and 5.

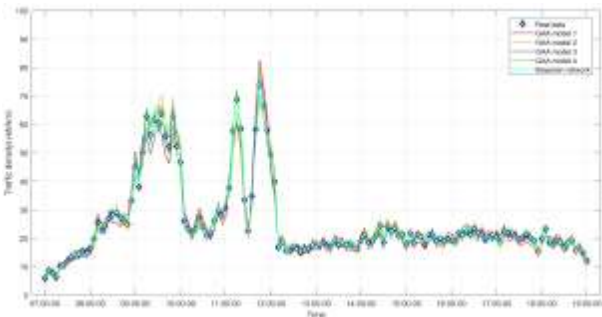

FIGURE 5(k) Predicted traffic density between loop detector 5 and 6.

**FIGURE 5 Model comparison based on the data collected for the I-5 freeway segment between 7:00AM and 7:00PM on October 10th.**

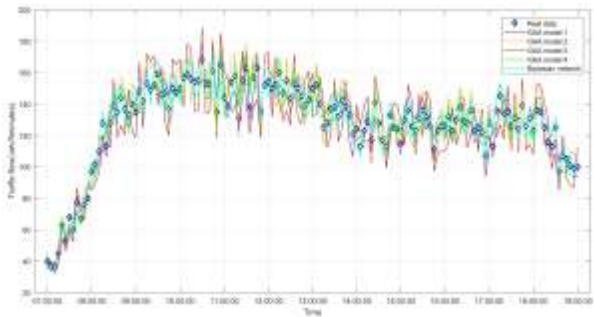

FIGURE 6(a) Predicted traffic flow for loop detector 1

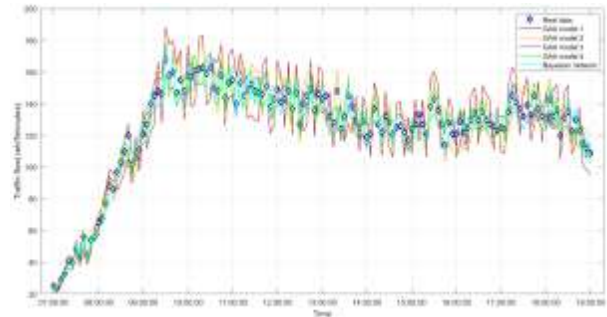

FIGURE 6(b) Predicted traffic flow for loop detector 2

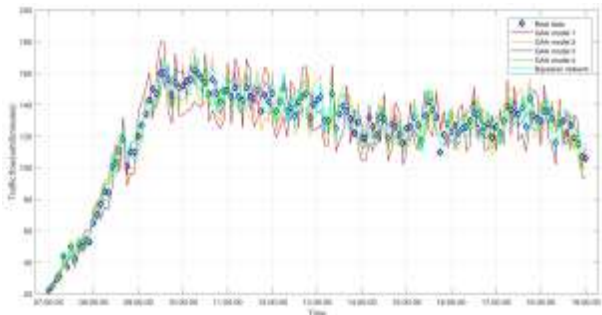

FIGURE 6(c) Predicted traffic flow for loop detector 3.

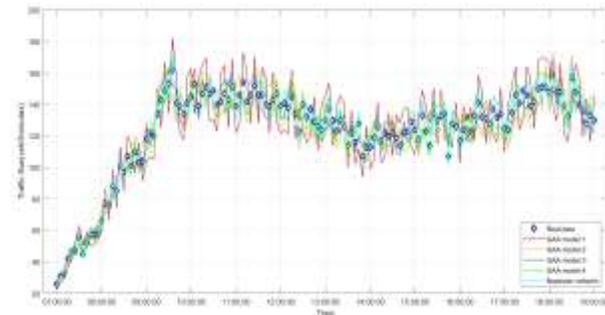

FIGURE 6(d) Predicted traffic flow for loop detector 4.

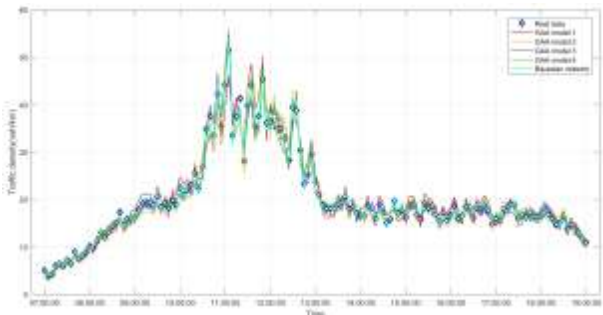

FIGURE 6(e) Predicted traffic density between loop detector 1 and 2.

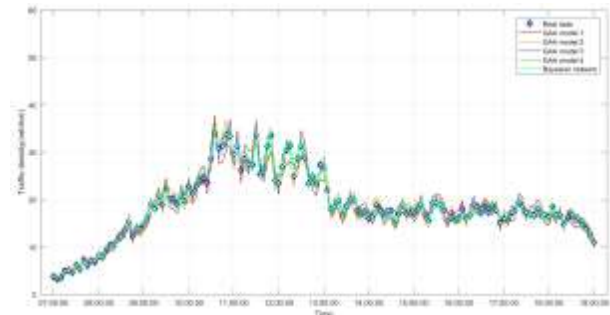

FIGURE 6(f) Predicted traffic density between loop detector 2 and 3.

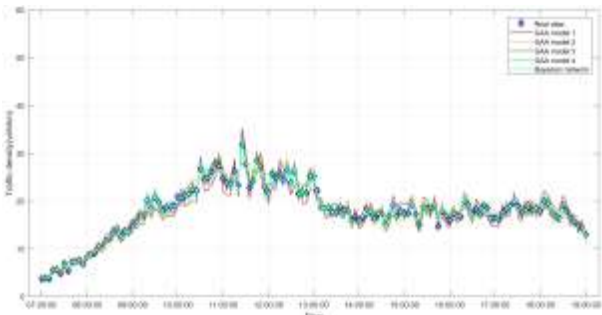

FIGURE 6(g) Predicted traffic density between loop detector 3 and 4.

**FIGURE 6 Model comparison based on the data collected for the CA-52 state highway segment between 7:00AM and 7:00PM on October 10th.**

To measure the effectiveness of the five models, the Mean Absolute Percentage Error (MAPE) and Mean Squared Error (MSE) are calculated. The MAPE and MSE obtained by the five models for traffic density estimation and traffic flow estimation are shown in Table 1.

## 4.1 Experiment on the data collected from Seattle

For traffic density estimation, the GAA without conservative loss performs almost as well as the GAA with conservative loss and perceptual loss, with only 0.08% larger MAPE. They both outperform the other two GAA models in terms of MAPE. While in terms of MSE, the GAA with perceptual loss and conservative loss outperforms the other three GAA models. In this case, the GAA with perceptual loss and conservative loss receive 2.84% lower MSE than GAA without conservative loss.

For traffic flow estimation, the GAA conservative loss and perceptual loss performs best among all GAA models in terms of both MAPE and MSE.

For traffic density estimation, the proposed GAA with conservative loss and perceptual loss outperforms the Bayesian network approach with 1.00% lower MAPE and 1.66 lower MSE. For traffic flow estimation, the GAA with conservative loss and perceptual loss performs nearly as well as the Bayesian network approach with only 0.07% larger MAPE. While the GAA with conservative loss and perceptual loss outperforms the Bayesian network approach with 0.81 lower MSE.

For traffic density estimation, the GAA without conservative loss outperforms the Bayesian network approach in terms of MAPE with 0.92% lower MAPE. In all other cases, the Bayesian network approach outperforms the other three GAA models.

## 4.2 Experiment on the data collected from San Diego

For traffic density estimation, the GAA with conservative loss and perceptual loss outperforms the other three GAA models in terms of both MAPE and MSE. For traffic flow estimation, the GAA without conservative loss performs almost as well as the GAA with conservative loss and perceptual loss, with only 0.07% larger MAPE. They both outperform the other two GAA models in terms of MAPE. While in terms of MSE, the GAA with perceptual loss and conservative loss outperforms the other three GAA models. In this case, the GAA with perceptual loss and conservative loss receives 3.54 lower MSE than the GAA without conservative loss.

For traffic density estimation, the proposed GAA with conservative loss and perceptual loss outperforms the Bayesian network approach with 0.95% lower MAPE and 1.04 lower MSE. The Bayesian network approach outperforms the GAA without conservative loss with 0.89% lower MAPE and 3.09 lower MSE. For traffic flow estimation, the performance of the Bayesian network approach, the GAA without conservative loss and the GAA with conservative loss and perceptual loss are close in terms of MAPE.The difference between the MAPE of the Bayesian network approach and the MAPE of the GAA without conservative loss is only 0.05%. The difference between the MAPE of the Bayesian network approach and the MAPE of the GAA with conservative loss and perceptual loss are only 0.02%. While the GAA with conservative loss and perceptual loss outperforms the Bayesian network approach with 0.85 lower MSE. In this case, the Bayesian network approach outperforms the GAA without conservative loss with 2.79 lower MSE.

In the first experiment, for traffic flow estimation, the difference between the GAA without conservative loss and the GAA without perceptual loss is marginal in terms of MAPE with only 0.08% MAPE difference. While in other cases, the GAA without conservative loss outperforms the GAA without perceptual loss. In the second experiment, the GAA without conservative loss outperforms the GAA without perceptual loss in traffic density estimation and traffic flow estimation in terms of both MAPE and MSE. This indicates that learning from the training data set brings greater improvements in traffic state estimation accuracy than learning from the traffic flow conservation law.

In both experiments, for both traffic density estimation and traffic flow estimation, the GAA without conservative loss and perceptual loss performs worst in terms of both MAPE and MSE.

Note that in both experiments, in traffic flow estimation, the difference between the Bayesian network approach and the GAA with conservative loss and perceptual loss is marginal in terms of MAPE. It seems that the GAA with conservative loss and perceptual loss has more advantages over the Bayesian network approach in traffic density estimation compared with that in traffic flow estimation.

**TABLE 1 Performance of different models in traffic state estimation**

| Experiment 1 | | | | |
|---|---|---|---|---|
| | Traffic density estimation | | Traffic flow estimation | |
| | MAPE | MSE | MAPE | MSE |
| GAA without perceutal loss and conservative loss | 11.47% | 125.62 | 12.36% | 110.55 |
| GAA without perceptual loss | 9.80% | 20.55 | 7.91% | 19.30 |
| GAA without conservative loss | 5.12% | 9.11 | 7.83% | 8.42 |
| GAA with perceptual loss and conservative loss | 5.04% | 6.27 | 5.26% | 5.95 |
| Bayesian network | 6.04% | 7.93 | 5.19% | 6.76 |
| Experiment 2 | | | | |
| | Traffic density estimation | | Traffic flow estimation | |
| | MAPE | MSE | MAPE | MSE |
| GAA without perceutal loss and conservative loss | 10.76% | 119.59 | 12.08% | 108.13 |
| GAA without perceptual loss | 9.07% | 19.88 | 9.16% | 25.64 |
| GAA without conservative loss | 6.42% | 10.23 | 6.02% | 7.92 |
| GAA with perceptual loss and conservative loss | 4.68% | 6.10 | 5.95% | 4.28 |
| Bayesian network | 5.53% | 7.14 | 5.97% | 5.13 |

## 5 DISCUSSIONS AND CONCLUSIONS

This study proposes a generative adversarial architecture for spatial-temporal traffic state estimation. The proposed GAA is compared with the existing Bayesian network approach on loop detector data collected from Seattle, Washington and that collected from San Diego, California. Experimental results indicate that as the GAA incorporates traffic flow theory, it can achieve higher accuracy in traffic state estimation than the Bayesian network approach. In addition, two useful findings can be summarized from the experiments: 1) adding the conservative loss can improve the accuracy of traffic state estimation, and adding the perceptual loss can lead to further improvement; 2) the GAA with conservative loss and perceptual loss is better at density and traffic flow estimation than the Bayesian network approach, and the magnitude of the benefit is stronger for density estimation.

The proposed GAA can be applied in real-time traffic prediction tasks, as well as in historical traffic state reconstruction. Further, artificial intelligent technology is shown to be promising in transportation area. Even though a GAA requires a large amount of training traffic data to be trained and quite a long time to achieve Nash equilibrium in the training procedure, these two shortcomings can be easily tackled with the aid of enriched data collected by various advanced sensors and advanced computation technologies.

Despite the promising results obtained from this study, traffic flow phenomena, such as breakdown and capacity drop, cannot be caught by the proposed GAA in our experiments. The difficulty of using artificial intelligent technology for predicting complicated traffic flow phenonmenon lies in two aspects. The first aspect is that it is difficult to understand how they can learn to predict breakdown and capacity drop from training data. Moreover, if there is no traffic flow breakdown events described in training data set, it is almost impossible for the proposed GAA to create knowledge about breakdown. Future research directions can work to modify the GAA and make it able to learn and predict more complicated traffic flow phenomena.

## ACKNOWLEDGMENTS

This research was supported in part by the National Natural Science Foundation of China (Grant No. 51329801) and the Shenzhen Science and Technology Planning Project (Grant No. GJHZ20150316154158400).